\documentstyle[multicol,epsf,pre,aps]{revtex}

\newcommand{\be}{\begin{eqnarray}}
\newcommand{\ee}{\end{eqnarray}}

\begin{document}
\date{\today}
\draft

\title{Boundary Layer Analysis of the 
Ridge Singularity in a Thin Plate.}

\author{Alexander E. Lobkovsky}
\address
{The James Franck Institute\\
The University of Chicago\\
5640 South Ellis Avenue
Chicago, Illinois 60637}
\maketitle

\begin{abstract}
Large deformations of thin elastic plates and shells present a 
formidable problem in continuum mechanics which is generally intractable
except by numerical methods.  Conventional approaches break
down in the limit of small plate thickness due to appearance of
discontinuities in the solution which require boundary layer 
treatment.  We examine a simple case of a plate bent by forces
exerted along its boundary so as to create a sharp crease in the
limit of infinitely small thickness.  We find a separable boundary 
layer solution of the von Karman plate equations which is valid along
the ridge line.  We confirm a scaling
argument \cite{fullerine} that the ridge possesses a characteristic
radius of curvature $R$ given by the thickness of the sheet $h$ 
and the length of the ridge $X$ {\it viz.} $R \sim h^{1/3} X^{2/3}$.
The elastic energy of the ridge scales as $E \sim \kappa (X/h)^{1/3}$
where $\kappa$ is the bending modulus of the sheet.  We determine the
dependence of these quantities on the dihedral angle of the ridge 
$\pi - 2\alpha$.
For all angles $R \sim \alpha^{-4/3}$ and $E \sim \alpha^{7/3}$.
The framework developed in this paper is suitable for determination of other 
properties of ridges such as their interaction or  
behavior under various types of loading.  We expect these results 
to have broad importance in describing forced crumpling
of thin sheets.
\end{abstract}

\begin{multicols}{2}
\section{INTRODUCTION}

Mechanical properties of a thin elastic sheet
undergoing large distortions is a rich problem
which is still largely unresolved despite great
effort \cite{postbuckling}.
The postbuckling behavior of thin-walled structures is
of importance in safety design and in development of energy 
absorbing structures \cite{crash,cushion}.
Some microscopic solid-like membranes can also be found in a
crumpled state.  Examples of such membranes include phospholipid bilayers
below the 2D freezing point \cite{vesicle}, networks of tropomyosin 
\cite{rabbit},
and monomolecular layers such as graphite oxide \cite{graphite} or molybdenum
disulfide \cite{rag}.  In an entirely different context, 
formation of mountain ranges is a result of strong deformation of
earth's tectonic plates which to an extent behave as elastic sheets 
\cite{geophys}.

Several types of distortion in elastic membranes have been recently
analyzed.  Thermal 
fluctuations of a flat membrane roughen the surface, but preserve its 
overall flat shape\cite{david}.  
The surface is thought to lose its flatness for sufficiently high temperature,
at the ``crumpling transition.''
By contrast, certain defects deliberately 
introduced into flat membrane destroy its flatness.  The energy associated 
with these distortions is well understood\cite{seung}.
Different distortions occur when a flat membrane is collapsed by an external
force.   The incipient deformation in response to a load and the associated
buckling  instabilities are classic subjects of continuum 
mechanics\cite{landau,yamaki}.  

Large deformations of thin plates and shells can be described by 
two coupled quadratic fourth order partial differential equations whose
explicit solution exists only in a handful of cases \cite{platebook}.
The main complication is, however, that the small parameter related
to the thickness of the shell usually multiplies the highest derivative
or the nonlinear term in the equations \cite{fung}.  This leads to
formation of boundary layers in the asymptotic limit of small
shell thickness.  An example in which such boundary layer appears
is a thin plate bent by torques applied at the boundary first
analyzed by Kelvin, Tait and Friedrichs\cite{friedrichs}.
Other well-known boundary layer
phenomena which include the Prandtl's boundary layer in a flow at large
Reynolds numbers \cite{prandtl} arise under similar conditions.

The novel character of the boundary layer problem considered in this
article is in its potential relevance to a broad range of crumpling
phenomena.  A number of empirical studies have investigated the 
statistical scaling properties of the crumpled state
\cite{gomes,abraham,kardar}.  Current theoretical understanding of this forced crumpling
is in its initial stages, however.  
It was argued in Ref.~\cite{science} that the structure of
a thin crumpled sheet can be thought of as a collection of flat facets
bound by a network of straight ridges which terminate at sharp
vertices.  These ridges have the same nature as the boundary 
layer arising in a simple geometry considered in this article.
The scaling properties of the boundary layer solution allow
one to make far-reaching conclusions about the morphology and
the elastic energy of a crumpled thin sheet.

In the next section we derive the equations which describe large
deflections of thin elastic plates and introduce the boundary
value problem which exhibits the novel boundary layer.  Scaling
properties of this boundary layer are examined in Section III.
A separation of variables Ansatz is introduced in Section IV.
The Ansatz implies that the local properties of the ridge solution
scale with the distance from
the boundary in a simple way.  In Section V we derive the scaling
of the separable solution with the dihedral angle of the ridge $\alpha$.
From here on we will refer to $\alpha$ as 
the ``dihedral angle'' even though it is a half of the difference of the
dihedral angle from $\pi$.  Confusion is unlikely to result
since this is the only way $\alpha$ is used in the paper.
Corroborating numerical evidence is presented in Section VI.
Finally, the implications of the thickness scaling for the crumpling
problem and the validity of the separable solution as well as future
work are discussed in Section VII.

\section{VON KARMAN PLATE EQUATIONS}

We start with a brief review of the theory of large
deflections of thin plates originally due to von Karman\cite{vonK_original}.
The equations are usually obtained within an expansion scheme of the
full three dimensional elasticity equations
with the thickness of the plate
as a small parameter.  Small strains and a linear
stress-strain relation are usually assumed\cite{justification}.  We however
prefer to derive the equations in a somewhat different way which is more
instructive while not as rigorous as the thickness expansion method.

When the thickness of the plate is small the in-plane stresses
are much greater than the normal stresses, so that
the dependence of the in-plane stresses on the normal coordinate
is simple \cite{landau}.  Then, all variables can be integrated
over the thickness of the plate and it can be 
treated as a two-dimensional surface.  The plate has a preferred
material coordinate system ${\bf x} \in D$ some open simply
connected domain in ${\bf R}^2$.
According to the Fundamental Theorem of Surfaces\cite{diff_geom}, 
to define a surface
uniquely up to an overall translation and rotation
in ${\bf R}^3$ it is sufficient to specify
two symmetric tensors in $D$: a metric tensor
$g_{\alpha\beta}$ and a curvature tensor $C_{\alpha\beta}$ satisfying
a set of relations which we will write down shortly.  To clarify the
meaning of these tensors we first 
note that the strain tensor $\gamma_{\alpha\beta}$
is defined as the deviation of the metric tensor from the identity

\be
g_{\alpha\beta}=\delta_{\alpha\beta}+2\gamma_{\alpha\beta}.
\ee

\noindent The sum of the eigenvalues of the strain tensor $\gamma_1+\gamma_2$
is the (2D) expansion (or compression) factor and their difference is the 
shear angle\cite{landau}.
The eigenvalues of the curvature tensor, on the other
hand, are the inverses of the two principal radii of curvature of the
surface.  Given the three-dimensional position $\vec{\bf r}(x_1,x_2)$ of the
material point $(x_1,x_2)$ the metric tensor and the curvature tensor are
given by\cite{diff_geom}

\be
g_{\alpha\beta} & = &
(\partial_\alpha \vec{\bf r}) \cdot (\partial_\beta \vec{\bf r})\\
C_{\alpha\beta} & = &
\hat{\bf n} \cdot (\partial_\alpha\partial_\beta \vec{\bf r}),
\ee

\noindent where $\partial_\alpha$ denotes differentiation with respect to
the material coordinate $x_\alpha$, and $\hat{\bf n}$ is the unit normal
to the surface.
To be absolutely rigorous one must distinguish upper and lower indices
but since we define all quantities to first nontrivial order
in the strain $\gamma_{\alpha\beta}$ and raising and lowering of the indices
is accomplished by applying the metric tensor $g_{\alpha\beta}$,
raising and lowering indices only affects higher order terms in
$\gamma_{\alpha\beta}$.

In order for the tensors $g_{\alpha\beta}$ and $C_{\alpha\beta}$ to
define a surface they must satisfy two relations involving
the Christoffel symbols $\Gamma_{\alpha\beta\mu}$ \cite{diff_geom}
which are defined in terms of the metric tensor.  When the
strains are small the expression is particularly simple:

\be
\Gamma_{\alpha\beta\mu} = -\partial_\mu \gamma_{\alpha\beta} +
\partial_\alpha \gamma_{\beta\mu} + \partial_\beta \gamma_{\alpha\mu}.
\ee

\noindent  
The first relation which $g_{\alpha\beta}$ and $C_{\alpha\beta}$
must satisfy in order to define a surface 
is the Gauss's {\it Theorema Egregium} \cite{diff_geom}.
It expresses the Gaussian curvature $K={\rm det\ }C_{\alpha\beta}$
which is the determinant of the
curvature tensor in terms of the Christoffel symbols.
In the case of small and slowly varying strain ({\it i.e.} 
$\partial^2\gamma \gg (\partial\gamma)^2$) the Gauss's theorem
reads

\be
K & = &\partial_2\Gamma_{211}
-\partial_2\Gamma_{212} + \Gamma_{11\rho}\Gamma_{\rho22} -
\Gamma_{12\rho}\Gamma_{\rho21}\nonumber\\
 & = &
\partial_\alpha\partial_\beta \gamma_{\alpha\beta} - 
\nabla^2{\rm tr}\ \gamma_{\alpha\beta},
\label{gauss}
\ee

\noindent where $\nabla^2 = \partial_\mu\partial_\mu$.  Summation
over repeated indices is implied.
Geometrically, the Gauss's equation captures the intuitive notion that
non-zero Gaussian curvature (the sheet curves in both directions)
must cause the sheet to strain.  Due to historical reasons
Eq.\ (\ref{gauss}) in this context is usually referred to as the first
von Karman equation.

The other set of relations usually termed as Codazzi-Mainardi
equations describes how the curvature tensor
behaves when transported around a closed curve.
Again, when the strains and consequently the Christoffel symbols
are small these relations are simple.
They say that if the curvature tensor is parallel 
transported around a closed curve, the change is of 
higher order in the strain\cite{diff_geom}

\be
\partial_\gamma C_{\alpha\beta} = \partial_\beta C_{\alpha\gamma}.
\label{codazzi}
\ee

\noindent These equations are a tensor analogy of the condition
satisfied by an irrotational vector field.
And just as an irrotational vector field can be written
as a gradient of a scalar, there exists a scalar function $f$ such that

\be
C_{\alpha\beta}=\partial_\alpha\partial_\beta f.
\label{def_f}
\ee

Equations Eq.\ (\ref{gauss}) and (\ref{codazzi}) ensure that
$C_{\alpha\beta}$ and $\gamma_{\alpha\beta}$ describe a physical
surface.  We now consider further conditions which assure that
each element of the surface is in mechanical equilibrium.
The forces and the torques acting on an infinitesimal element of the
surface $\delta x\delta y$ can be found from the
the stress tensor $\sigma_{\alpha\beta}$ and
the torques $M_{\alpha\beta}$ \cite{platebook} as shown
in Fig.~1a and 1b.  The force tensors are related to
the deformation tensors $\gamma_{\alpha\beta}$ and $C_{\alpha\beta}$
by the constitutive relations for an elastic body.
For sufficiently small strains,

\be
\sigma_{\alpha\beta}={Y \over 1-\nu^2}\left[\gamma_{\alpha\beta} +
\nu \epsilon_{\alpha\rho}\epsilon_{\beta\tau}\gamma_{\rho\tau}\right],
\label{stress}
\ee

\noindent where $Y$ is the Young's modulus and $\nu$ is the Poisson
ratio of the elastic material \cite{landau}.  Here
$\epsilon_{\alpha\beta}$ is the two dimensional
antisymmetric tensor. There is a similar relation
for the torques per unit length

\be
M_{\alpha\beta} = \kappa \left[C_{\alpha\beta} + 
\nu\epsilon_{\alpha\rho}\epsilon_{\beta\tau}C_{\rho\tau}\right]
\label{torque}
\ee

\noindent which comes from considering the stress distribution throughout
the thickness of the plate $h$\cite{platebook}.
Here $\kappa=Yh^3/(12(1-\nu^2))$ 
is the bending rigidity of the plate\cite{landau}.
The in-plane force equilibrium can be written down by inspection (see
Fig.~1a).  It reads

\be
\partial_\alpha \sigma_{\alpha\beta} = 0.
\ee

\noindent This condition allows one to write the stresses in terms of a
scalar potential $\chi$ traditionally called the force function:

\be
\sigma_{\alpha\beta} = \epsilon_{\alpha\mu}\epsilon_{\beta\nu}
\partial_\mu \partial_\nu \chi.
\ee

\noindent This relation is a tensor analog of the divergenceless vector
field expressed as a curl of a vector potential.

\begin{figure}
\centerline{\epsfxsize=8.5cm \epsfbox{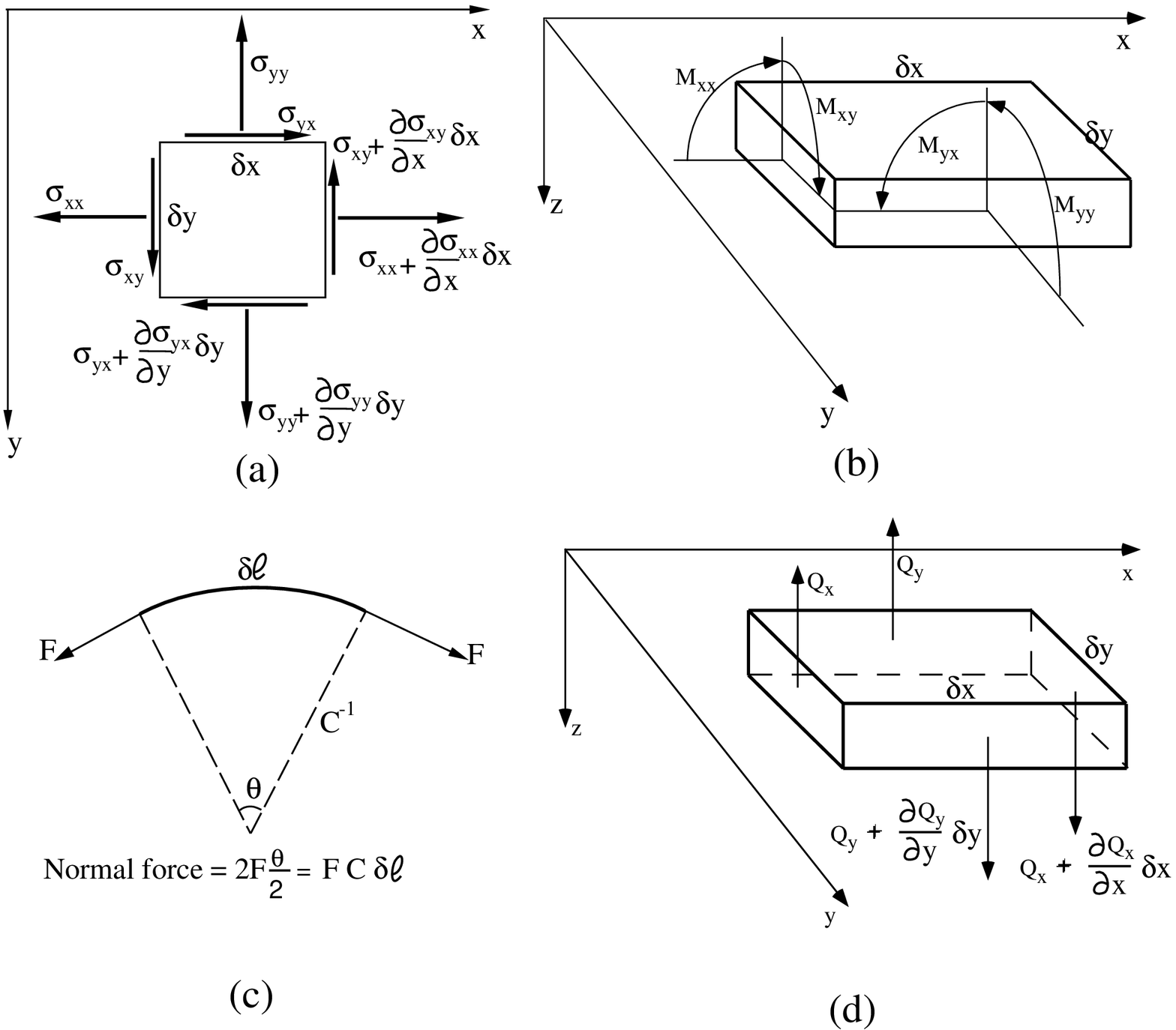}}
{FIG. 1. (a) In-plane stresses (per unit length) acting on a
small element of an elastic sheet,  (b) Torques per unit
length acting on the same element,  (c) One-dimensional
derivation of the normal force on a curved  element $\delta l$
due to in-plane forces $F$,  (d) Normal shear stresses acting over the
sides of the elementary area of the sheet.}
\end{figure}

The resultant normal force on the element of the surface can come
from three sources.  First, is the external distributed load which we for the
time being will ignore.  Second, the normal components of the
in-plane tensions that act on the element are non-zero due to
the curvature of the surface.  It is easy to become convinced
(see Fig.~1c for a one dimensional version of the
derivation) that that normal force per unit area 
due to the in-plane tension is $h \sigma_{\alpha\beta}
C_{\alpha\beta}$\cite{platebook}.
Third, changing torques in the sheet produce a normal force.
To see this, we define the normal shear stresses $Q_\alpha$ acting over the
sides of our surface element as shown on Fig.~1d.
The normal force on the element due to the variation of these
shear stresses is clearly $\partial_\alpha Q_\alpha \delta x\delta y$.
The normal shear stresses $Q_\alpha$ can be found by balancing the moments
about the $x$- and $y$-axes of all the forces acting on the element
which gives $Q_\alpha = \partial_\beta M_{\alpha\beta}$.
Collecting the normal forces we obtain the second
von Karman equation

\be
\partial_\alpha \partial_\beta M_{\alpha\beta}=
h\sigma_{\alpha\beta} C_{\alpha\beta}.
\label{normal}
\ee

The more familiar form of the von Karman equations emerges when
one substitutes the potentials $f$ and $\chi$ into Eqs.\ (\ref{gauss}) and 
(\ref{normal})
using Eq.\ (\ref{stress}), Eq.\ (\ref{torque}), 
and the definition of the potentials.  They read

\begin{equation}
\left\{
\begin{array}{rcl}
\kappa \nabla^4 f & = & \ [\chi,f] \\
\displaystyle{{1 \over Y}} \nabla^4 \chi & = & -\displaystyle{{1 \over 2}}[f,f],
\end{array}
\right.
\label{vonK}
\end{equation}

\noindent where we have defined 
\be
[a,b] & \equiv &  \epsilon_{\alpha\mu}\epsilon_{\beta\nu}
(\partial_\alpha\partial_\beta a)
(\partial_\mu\partial_\nu b) \nonumber\\
& = &
{\partial^2 a \over \partial x^2} 
{\partial^2 b \over \partial y^2} +
{\partial^2 a \over \partial y^2}
{\partial^2 b \over \partial x^2} -
2{\partial^2 a \over \partial x \partial y}
{\partial^2 b \over \partial x \partial y}.
\ee

\noindent Notice that $[f,f]$ is twice the Gaussian curvature $K$.

In principle, the strains $\gamma_{\alpha\beta}$ and
the curvatures $C_{\alpha\beta}$ obtained from the von Karman equations 
as a function of the material
coordinates define the surface uniquely (up to position in ${\bf R}^3$).
The problem of finding the shape of the surface from the
strains and the curvatures 
is however highly nonlinear and intractable in general.
On can make progress in a limited class of deformations in which the
normal to the surface does not change much.  
In that case the so called Monge coordinates are appropriate.
The undeformed plate is located in the $x-y$ plane so that upon
deformation the point originally at
$(x_1, x_2, 0)$ moves to $(x_1 + u_1, x_2 + u_2, w)$ in the 3-dimensional
embedding space where $u_\alpha$ and $w$ are functions of $x_\alpha$.
If the derivatives of $u_\alpha$ and $w$ are small everywhere, only the
lowest non-trivial order terms in those derivatives can be kept in 
the expressions for the strains and the curvatures.  We obtain\cite{landau}

\be
\gamma_{\alpha\beta} = {1 \over 2}\left(\partial_\alpha u_\beta +
	\partial_\beta u_\alpha + 
	\partial_\alpha w\partial_\beta w\right)
\ee

\noindent and

\be
C_{\alpha\beta} = \partial_\alpha\partial_\beta w.
\ee

\noindent
It is clear that the normal displacement $w$ in this case is precisely the
potential function $f$ defined in Eq.\ (\ref{def_f}).

To complete the description of the deformation of the thin elastic plate
one needs to be able to calculate the elastic energy stored in the 
sheet and specify the boundary conditions
on the functions $f$ and $\chi$.
We begin by considering the work done on the
small element of the surface $\delta x \delta y$ by the surrounding
parts of the plate when the strain in the element changes by 
$\delta \gamma_{\alpha\beta}$.  That work is
$h\sigma_{\alpha\beta}\delta\gamma_{\alpha\beta}\delta x \delta y$
\cite{platebook}.
The stretching energy $E_s$ in the plate element is found by
integrating the strain from $0$ to its value $\gamma_{\alpha\beta}$
while keeping in mind that the stresses are proportional to the
strains which introduces a factor of $1/2$.  Thus the total stretching
energy in the plate is given by

\be
E_s = {h \over 2} \int dxdy\ \sigma_{\alpha\beta}\gamma_{\alpha\beta}.
\ee

\noindent One can similarly show that the work done by the torques
$M_{\alpha\beta}$ in bending a surface element is
${1 \over 2}M_{\alpha\beta}C_{\alpha\beta}\delta x \delta y$ so that
the total bending energy $E_b$ in the plate is \cite{platebook}

\be
E_b = {1 \over 2} \int dxdy\ M_{\alpha\beta}C_{\alpha\beta}.
\ee

\noindent Expressing the strains, stresses, torques and curvatures
in terms of the potentials $\chi$ and $f$ we get after some algebra

\be
E_s & =  \displaystyle{h \over 2Y} & \int dx dy\ 
	\left[{\rm tr}(\partial_\alpha\partial_\beta \chi)\right]^2 -
	2(1+\nu) {\rm det}(\partial_\alpha\partial_\beta \chi)
\label{E_s} \\
E_b & =  \displaystyle{\kappa \over 2} & \int dx dy\ 
	\left[{\rm tr}(\partial_\alpha\partial_\beta f)\right]^2 -
	2(1-\nu) {\rm det}(\partial_\alpha\partial_\beta f).
\label{E_b}
\ee

\noindent One can show that the conformation ${\bf r}(x_1, x_2)$
of the sheet which
is a solution of the von Karman equations minimizes the elastic 
energy while satisfying the boundary conditions\cite{justification}.
We note in passing that many derivations of the von Karman
equations (see {\it e.~g.}~Ref.~\cite{seung}) start with writing down
these energies in an add-hoc way and then taking a variational
derivative with respect to the sheet shape.

We are left to address the question of the
boundary conditions. As pointed out in Ref. \cite{justification},
specification of the
boundary conditions for the von Karman equations is tricky
and is rarely done in a completely rigorous or even correct manner.
The main problem arises when a fixed shape of the boundary in 
the embedding space is prescribed.  Since the equations are
in terms of the material coordinates, the knowledge of the
shape in the embedding space requires solving the equations
first.  Therefore, a boundary condition that makes
reference to the embedding space is intractable.
Only in the case when the material
coordinates can be replaced with the Monge coordinates
mentioned above can one prescribe a fixed shape the boundary.
In other cases the best one can do is to specify the curvature
tensor and the strain tensor at the boundary of the
{\it material coordinates domain.}
One can imagine pulling or pushing on the edge of the plate
which amounts to specifying the stresses at the edges.
Torques $M_{\alpha\beta}$ and normal forces $Q_\beta$ can also 
be applied at the edge.

\begin{figure}
\centerline{\epsfxsize=8cm \epsfbox{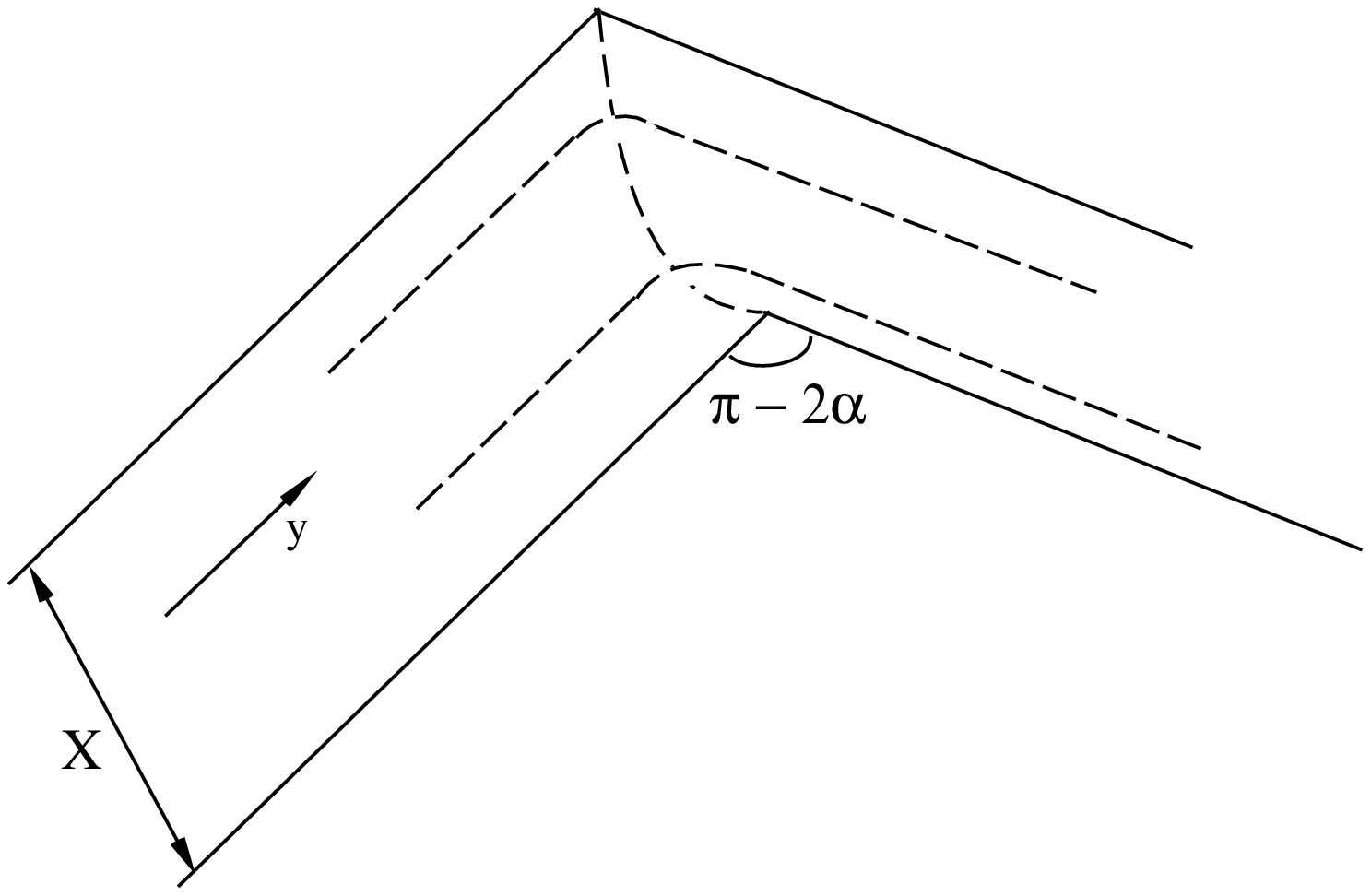}}
{FIG. 2. Long strip of width $X$ bent through
a dihedral angle $\alpha$.}
\end{figure}

We proceed to define the boundary value problem which exhibits
the ridge singularity.
We consider a strip $(x, y) \in (-X/2, X/2)\times
(-\infty, \infty)$ of uniform thickness
$h$ made of isotropic homogeneous material
with Young's modulus $Y$ and Poisson ratio $\nu$.  
Normal forces are applied to the edge so as to bend the strip
by an angle $\pi - 2\alpha$ (see Fig.~2).
The membrane stresses $\sigma_{\alpha\beta}$ as well as the
torques $M_{\alpha\beta}$ vanish at the boundary (except for the
singular point $y=0$).  In terms of the
potentials $f$ and $\chi$ it means that

\be
\partial_\alpha\partial_\beta f = 
\partial_\alpha\partial_\beta \chi = 0 \hspace {1em} 
{\rm at}\hspace {1em}x=\pm {X \over 2}.
\label{boundary}
\ee

\noindent The condition that the strip is bent translates into
specifying the curvature tensor at the boundary.  We assume that
the boundary is a geodesic of the surface which is reasonable
in the small strain limit.  In the direction
along the boundary the curvature is zero except for a sharp peak at the
origin $y = 0$ where the strip is bent.  The width of the
bent region must be of the order of the thickness of the plate
and the curvature is of the order of the inverse thickness.
It is convenient then to set the curvature along
the boundary to a $\delta$-function since the
all the length scales in the problem are much larger than $h$.
This leads to a particularly simple condition on $f$ at the
boundary 

\be
f(\pm X/2, y) = \alpha |y|
\label{boundary1}
\ee

\noindent up to an arbitrary linear
function of $x_\alpha$.  To see that the coefficient $\alpha$
is identical to the bending angle introduced above consider
the following integral along the boundary (or any other geodesic which
approaches $x = const$)

\be
{\partial f \over \partial y}(\infty) = 
\int_0^\infty dy {\partial^2 f \over \partial y^2} = 
\int_0^\infty dy 
\left\vert{\partial \hat{\bf n} \over \partial y}\right\vert
=\alpha,
\ee

\noindent where we have used the definition of the curvature
tensor $C_{yy} = \partial_y \hat{\bf n} \cdot \hat{\bf t}_y$ 
and the fact that the integral is taken along a geodesic.  Here 
$\hat{\bf n}$ is the unit normal and
$\hat{\bf t}_y$ is tangent vector to the surface in the $y$-direction
\cite{diff_geom}.  The geometric meaning of this integral is the length
of the contour which is inscribed by the end of the 
normal vector $\hat{\bf n}$
on the unit sphere as $\hat{\bf n}$ is transported along the
geodesic.  Using this argument and a reasonable belief that
the solution far away from the ridge approaches that of an 
unperturbed flat strip on can deduce that the derivative of the
potential $f$ along the transverse direction $y$ does not
depend on $x$ far away from the ridge.  This fact will 
later allow us to gain insight into the nature of the solution.

The physical reason one expects non-trivial singular behavior
in the small thickness limit is the interplay between
the bending and the stretching energies.
The key observation is that the bending modulus $\kappa$ vanishes 
faster with the thickness of the plate (as $h^3$) than the stretching modulus
$Yh$ \cite{landau} so that thin plates bend in preference to stretching.
Therefore, in the limit of zero thickness, the bent plate develops a 
sharp crease at which the solution is irregular 
(the curvature in the $y$-direction
diverges).  At a finite thickness, as shown in ref.~\cite{fullerine},
the crease softens to a characteristic width governed by the
competition of the bending and the stretching energies 
in the region of the crease.

\section{BOUNDARY LAYER SCALING}

We first put the von Karman equations (\ref{vonK}) in the non-dimensional
form by defining

\be
\bar{\chi} = h {\chi \over \kappa}, \hspace{1em}
\bar{f} = {f \over X}, \hspace{1em}
\bar{x} = {x \over X} \hspace{1em} {\rm and} \hspace{1em}
\bar{y} = {y \over X}
\ee

\noindent to obtain

\begin{equation}
\left\{
\begin{array}{rcl}
\nabla^4 \bar{f} & = & [\bar{\chi},\bar{f}] \\
\lambda^2 \nabla^4 \bar{\chi} & = & 
-\displaystyle{{1 \over 2}}[\bar{f},\bar{f}]
\end{array}
\right.
\end{equation}

\noindent with the small dimensionless geometry parameter

\be
\lambda \equiv {\sqrt{\kappa/Y} \over X} = 
\left({h \over X}\right){1 \over \sqrt{12(1-\nu^2)}}
\ee

\noindent multiplying the highest derivative.
The solution to the so-called reduced problem defined by $\lambda=0$ 
can be readily obtained.  There is no cost for bending 
and thus the strip froms a sharp crease $f(x, y)=\alpha|y|$.
The properties of the boundary layer which corrects
the discontinuity in the reduced solution
be found by the rescaling

\be
\tilde f = \lambda^\beta \bar{f}, \hspace{1em} 
\tilde\chi = \lambda^\delta \bar{\chi}, \hspace{1em}
\tilde x = \lambda^0 \bar{x}, \hspace{1em}
\tilde y = \lambda^\beta \bar{y}.
\ee

\noindent Note that $\bar{x}$ remains unchanged by the
rescaling transformation and $\bar{f}$ and $\bar{y}$
are rescaled by the
same factor to satisfy the boundary condition (\ref{boundary1}.
The rescaled equations read:

\vspace{2mm}
\begin{equation}
\left\{
\begin{array}{rcl}
\lambda^{-\beta}
\left[\displaystyle{{\partial^4 \tilde f \over \partial \tilde x^4}} + 
2\lambda^{2\beta} 
\displaystyle{
{\partial^4 \tilde f \over \partial^2 \tilde x \partial^2 \tilde y}} +
\lambda^{4\beta} \displaystyle{
{\partial^4 \tilde f \over \partial^4 \tilde y}}\right] & = &
\lambda^{\beta - \delta}[\tilde\chi,\tilde f] \\[3mm]
\lambda^{2-\delta} \left[\displaystyle{
{\partial^4 \tilde\chi \over \partial \tilde x^4}} + 
2\lambda^{2\beta} \displaystyle{
{\partial^4 \tilde\chi \over \partial^2 \tilde x \partial^2 \tilde y}} +
\lambda^{4\beta} \displaystyle{
{\partial^4 \tilde\chi \over \partial^4 \tilde y}}\right] & = &
\lambda^0\displaystyle{{1 \over 2}}[\tilde f,\tilde f].
\end{array}
\right.
\label{scaling}
\end{equation}
\vspace{2mm}

\noindent  The dominant terms in the $\lambda \rightarrow 0$ limit
must be of the same order on either side of the equations.  This leads
unambiguously to $\beta < 0$ which is in agreement with the intuitive
guess that the width of the boundary layer must go to zero as the thickness 
of the sheet vanishes.  Balance of the dominant terms gives

\be
\beta = -{1 \over 3} \hspace{1em} {\rm and} \hspace{1em} \delta = {2 \over 3},
\ee

\noindent in agreement with the scaling argument of Witten and Li.
With the knowledge of the scaling behavior of the functions $f$ and
$\chi$ on can obtain the small thickness behavior of other
quantities such as, for example, the transverse ridge curvature 
$C_{yy} = \partial^2 f / \partial y^2$
and the midridge longitudinal strain
$\gamma_{xx} =(1/Y)(\sigma_{xx} - \nu \sigma_{yy}) 
\simeq (1/Y) \partial^2 \chi / \partial y^2$ where the expressions
are evaluated at $x=0$ and $y=0$.  The leading order behavior
of the transverse stress
$\sigma_{yy} = \partial^2 \chi/\partial x^2$
is of higher order in $\lambda$ and it thus can be
ignored in the expression for $\gamma_{xx}$.  Substituting
the rescaled quantities we find that

\be
C_{yy} = {1 \over X} \lambda^{-1/3} 
{\partial^2 \tilde f \over \partial \tilde y^2}
\hspace{1em} {\rm and} \hspace{1em}
\gamma_{xx} = 
\lambda^{2/3} {\partial^2 \tilde \chi \over \partial \tilde y^2}.
\ee

\noindent These expressions give the asymptotic behavior of $C_{yy}$ 
and $\gamma_{xx}$ since the rescaled quantities do not depend on 
$\lambda$ in the small $\lambda$ limit.  

Note that the width of the
boundary layer has the same leading order behavior as
the radius of curvature of the plate at the center of the ridge
and as the ``sag'' which is the vertical deflection of the
ridge shape from that of perfectly sharp crease.
This deflection is given by $f(0,0)$ in the small dihedral angle limit
$\alpha \ll 1$.  The sag can be found for a general 
dihedral angle $\alpha$ from the following argument.
Due to the $x \rightarrow -x$ symmetry of the solution
the ridge line $y=0$ is a geodesic of the surface.
Therefore we can relate the vertical
deflection of the sheet $w(x)$ along the ridge
to the curvature $C_{xx} = \partial^2 f/\partial x^2$.
To lowest order in the strain we obtain

\be
{w'' \over \sqrt{1-(w')^2}} = C_{xx} = {\partial^2 f \over \partial x^2}.
\ee

\noindent Since $C_{xx}$ vanishes in the limit of small thickness,
the leading term in the $\lambda^{2/3}$ expansion of $w''(x)$ scales with
the same power of $\lambda$ as $\partial^2 f/\partial x^2$.
Thus the ridge sag $w(0)$ is of the same order as the ridge width
and as the midridge transverse radius of curvature.

We can also find asymptotic behavior of the bending
and stretching energies of the sheet.
We first note that in the expressions for the energies 
Eqs.\ (\ref{E_s}) and (\ref{E_b})
the terms involving ${\rm det}(\partial_\alpha\partial_\beta f)$ and
${\rm det}(\partial_\alpha\partial_\beta \chi)$ can be expressed
as an integral over the boundary\cite{landau}.
These integrals vanish identically for the boundary conditions
considered here.  Substituting the rescaled quantities which remain 
finite as $\lambda \rightarrow 0$ into Eqs.\ (\ref{E_s}) and (\ref{E_b})
we obtain

\be
E_b & = &\kappa \lambda^{-1/3} \int d\tilde x d\tilde y
\left({\partial^2\tilde f \over \partial \tilde y^2}\right)^2
\\
E_s & = &\kappa \lambda^{-1/3} \int d\tilde x d\tilde y
\left({\partial^2\tilde \chi \over \partial \tilde y^2}\right)^2,
\label{scale_ener}
\ee

\noindent in agreement with the energy scaling argument of Witten and Li
\cite{fullerine}.  For a fixed thickness, these energies grow qualitatively
slower (as $X^{1/3}$) than the energy of a sharp crease of size $X$
which grows linearly  with $X$.

\section{A SOLUTION TO THE REDUCED EQUATIONS}

The solution to the rescaled equations (\ref{E_s}) can be sought 
as an expansion in powers of $\lambda^{2/3}$,

\begin{equation}
\left\{
\begin{array}{rcl}
\tilde f & = & f_0 + \lambda^{2/3} f_1 + \lambda^{4/3} f_2 + \ldots \\[1mm]
\tilde\chi & = & 
\chi_0  + \lambda^{2/3} \chi_1 + \lambda^{4/3} \chi_2 + \ldots.
\end{array}
\right.
\label{expansion}
\end{equation}

\noindent Plugging the series into Eq.\ (\ref{scaling}) 
and matching the coefficients
of powers of $\lambda^{2/3}$ we can obtain equations for all orders
in the expansion Eq.\ (\ref{expansion}).  We get at zeroth order,

\begin{equation}
\left\{
\begin{array}{rcl}
\displaystyle{{\partial^4 f_0 \over \partial \tilde y^4}}
 & = & [\chi_0, f_0] \\[2mm]
\displaystyle{{\partial^4 \chi_0 \over \partial \tilde y^4}}
 & = & -\displaystyle{{1 \over 2}}[f_0, f_0].
\end{array}
\right.
\label{zeroth}
\end{equation}

\noindent We note in passing the equations for $f_1$ and $\chi_1$.  They read

\begin{equation}
\left\{
\begin{array}{rcl}
\displaystyle{{\partial^4 f_1 \over \partial \tilde y^4}} + 2
\displaystyle{{\partial^4 f_0 \over \partial \tilde x^2 \partial \tilde y^2}}
& = & [\chi_0,f_1]+[\chi_1,f_0] \\[3mm]
\displaystyle{{\partial^4 \chi_1 \over \partial \tilde y^4}} + 2
\displaystyle{{\partial^4 \chi_0 \over\partial\tilde x^2 \partial\tilde y^2}}
& = & -[f_0,f_1].
\end{array}
\right.
\end{equation}

\noindent At this point we must draw the readers' attention
to the fact that while $\tilde f$
and $\tilde \chi$ satisfy the boundary conditions Eq.\ (\ref{boundary}) or
Eq.\ (\ref{boundary1}) there is no reason to expect that $f_0$ and $\chi_0$
do.  In other words, it is likely that the expansion in powers
of $\lambda^{2/3}$ does not 
converge uniformly, so that for a fixed $\lambda$ higher orders
in the expansion Eq.\ (\ref{expansion}) become increasingly
important as the boundary
is approached.  A solution of the zeroth order equations (\ref{zeroth}) might
therefore be a good approximation to $\tilde f$ and $\tilde \chi$
only in a restricted area around the ridge away from the boundaries.

To help solve equations (\ref{zeroth}) we observe that the boundary conditions
do not introduce another length scale to the problem.  Therefore,
the transverse profile of the ridge ought to {\it scale} with the
distance from its midpoint $\tilde x = 0$.  We must find a functional
form of $f_0$ and $\chi_0$ such that the different $\tilde x = const$ slices 
are related to each other by a scale factor $q(\tilde x)$
depending on $\tilde x$ only.
By assuming scaling in $\tilde x$ we hope to decouple it from the new
transverse variable $\xi = \tilde y/q(\tilde x)$.  We assume that

\be
f_0(\tilde x, \xi) = q^\mu p_1(\xi), \hspace{2em}
\chi_0(\tilde x, \xi) = q^\eta p_2(\xi).
\label{Ansatz}
\ee
 
\noindent
As we show in the appendix,
variables separate for only one choice of $\eta = \mu = 1$
which means that the width of the ridge solution scales with the
distance from the vertex in the same way as the ridge sag.
As mentioned above, the boundary conditions on $f$
require that its transverse derivative $\partial f/\partial y$
approach a constant independent of $x$.  If a similar condition is to hold
for $f_0$ then inescapably $\mu=1$.
The validity of this scaling hypothesis needs to be corroborated by
some other means.
In Section VI we present numerical evidence supporting this scaling
Ansatz, here we only note that this evidence is convincing.

Substitution of the Ansatz Eq.\ (\ref{Ansatz}) into the zeroth order
equations (\ref{zeroth}) gives

\begin{equation}
\left\{
\begin{array}{rcl}
p_1'''' & = & q''q^2\left[p_1''p_2 + p_2'' p_1 - \xi(p_1'' p_2' +
p_2'' p_1')\right] \\[1mm]
p_2'''' & = & -q''q^2\left[p_1''p_1 - \xi p_1''p_1'\right].
\end{array}
\right.
\label{tran_p}
\end{equation}

\noindent Separation of variables provides the equation for the
scale factor $q''q^2=A$, where $A$ is some constant.
This equation, together with the condition that $q(x)$ be even
can be solved by a substitution $g(q) = q'(\tilde x)$ to yield,

\be
\tilde x={1\over \pi}\left[{\pi \over 2} - 
\arcsin\sqrt{{q(\tilde x) \over q(0)}} +
\sqrt{{q(\tilde x) \over q(0)}\left(1-{q(\tilde x) \over q(0)}\right)}\right]
\label{scale_factor}
\ee

\noindent where the separation of variable constant is related
to $q(0)$ via $A = -\pi^2 q^3(0)/2$.
The scale factor $q(\tilde x)$
has a singular derivative near the vertices, since for 
$\tilde x = 1/2 - \epsilon$ 

\be
q(\epsilon) \simeq \left({3\pi \over 2}\right)^{2/3} \epsilon^{2/3},
\ee

\noindent which reinforces the suspicion that the series expansion (\ref{expansion}) does
not converge uniformly.  

Therefore to test the predicted functional
form of the scale factor
one ought to look a local property of the numerical solution
away from the boundary and find its limit as $\lambda \rightarrow 0$.
We discuss such a solution in Section VI.
Numerically one has access to only a limited
range of thickness to size ratios. ($\lambda$ cannot be too small
otherwise lattice effects become significant.)  Nevertheless,
the Taylor expansion coefficients 
of the scale function $q(\tilde x)$ around $\tilde x = 0$
can be successfully extracted from the numerics.
Since there is one adjustable parameter $q(0)$ we must
look at the first two coefficients in the expansion
$q(\tilde x) = q(0)(1 + b_2 \tilde x^2 + b_4 \tilde x^4 + \ldots)$.
Expanding Eq.\ (\ref{scale_factor}) we find $b_2 = \pi^2/4 \simeq 2.47$
and $b_4 = \pi^4/48 \simeq 2.03$, whereas the numerics gives
$b_2 = 2.53\pm 0.04$ and $b_4 = 2.09 \pm 0.07$ where the errors
are obtained from the scatter of the data.

The transverse equations (\ref{tran_p}) can be integrated once by defining

\be
\phi_\beta(\xi) = p_\beta(\xi) - \xi p_\beta'(\xi) 
\hspace{2em} {\rm for} \hspace{1em}  \beta=1,2
\ee

\noindent We obtain

\begin{equation}
\left\{
\begin{array}{rcl}
\phi_1''-\displaystyle{{2 \over \xi}}\phi_1'& = & A \phi_1 \phi_2 + D_1\\[2mm]
\phi_2''-\displaystyle{{2 \over \xi}}\phi_2'& = & 
-\displaystyle{{1 \over 2}}A \phi_1^2 + D_2.
\end{array}
\right.
\label{tran_phi}
\end{equation}

\noindent Where $D_\beta$ are the integration constants allowing for
a non zero asymptote of the functions $\phi_\beta$. The finite
limit of these functions as $\xi \rightarrow \infty$
would imply a finite stress and curvature far away from the ridge.
However, since we expect the series Eq.\ (\ref{expansion}) to 
converge non-uniformly,
the large $\xi$ behavior of $\phi_\beta$ is likely to be unphysical.  
The functions $p_\alpha$ can then be readily found from 

\be
p_\beta(\xi) = 
G_\beta \xi + \xi \int_\xi^\infty d\zeta {\phi_\beta(\zeta) \over \zeta^2},
\label{p_from_phi}
\ee

\noindent where $G_\beta$ are integration constants.
A further analysis of the transverse equations
will be done elsewhere.  Here we only remark that
the simplicity and the apparent symmetry of the Eqs.\ (\ref{tran_phi})
must point to some physical symmetry in the problem
which forces this behavior to occur.
Perhaps further light can be shed onto the physical 
meaning of the functions $\phi_\beta$ by writing
down the expression for the elastic energy of the
sheet in terms of the separable solution.
The leading term in the $\lambda^{2/3}$-expansion of the total
energy as in Eq.\ (\ref{scale_ener}) reads

\be
E & = & \int d\tilde x d\tilde y {1\over q^2}
\left((p_1'')^2+(p_2'')^2\right)\nonumber\\
 & = &\int {d\tilde x \over q(\tilde x)}
\int {d\xi \over \xi^2}\left[\left({d\phi_1 \over d\xi}\right)^2
+ \left({d\phi_2 \over d\xi}\right)^2\right].
\label{ener_phi}
\ee

We can also make contact with the boundary condition $\tilde f \rightarrow
\alpha |\tilde y|$ as $\tilde y \rightarrow \infty$.
If the integral in Eq.\ (\ref{p_from_phi}) converges, this boundary condition
gives $G_1 = \alpha$.  Now one can use the $y\rightarrow -y$ symmetry
of the problem and sufficient smoothness of functions involved to
conclude that $p_1'(0)=0$ which leads to

\be
\int_0^\infty {d\zeta \over \zeta}
{d\phi_1(\zeta) \over d\zeta} = -\alpha.
\label{bound_alpha}
\ee

\section{Dihedral Angle Scaling of the Separable Solution}

Even without having solved the transverse equations
one can deduce the behavior of the solution for different
dihedral angles $\alpha$ just from the form of the
Eqs.\ (\ref{tran_phi}).
We begin by noting that there exists a two-parameter
family of transformations which produce new solutions.
For example if $\vec\phi(\xi)$ is a two-component 
solution, then 
$\vec\psi(\xi) = S_1\vec\phi(S_2\xi)$
is also a solution of the equations (\ref{tran_phi}) but with a different
separation of variable
constant $A'=S_2^2/S_1 A$.  We must allow for variation of $A$ since it will
undoubtedly depend on the dihedral angle.
Using the boundary condition on $\phi_1$ Eq.\ (\ref{bound_alpha}) we can
find the corresponding dihedral angle $\alpha' = S_1S_2\alpha$.
With this condition, there is a one-parameter
family of scale transformations which produce a solution
for the dihedral angle $\alpha'$ given that for the 
dihedral angle $\alpha$.
Let us fix a reference solution $\vec\phi_0$ with $\alpha=1$ so that all
other solutions are labeled by the scale factors
$S_1$ and $S_2$.  We can now find how various
quantities of interest are affected by the scale transformation.
We start with the simplest cases

\begin{equation}
\begin{array}{rcl}
\alpha & \sim & S_1S_2  \\
p_1 \sim p_2 & \sim &S_1 \\
q \sim  A^{1/3} & \sim & S_1^{-1/3}S_2^{2/3}.
\end{array}
\end{equation}

\noindent The relevant quantities such as the ridge sag
$qp_1$,
the transverse curvature in the middle of the ridge
$(1/q)p_1''$ both evaluated at $\tilde x = \xi = 0$
and the elastic energy Eq.\ (\ref{ener_phi}) all turn out to depend only on the
product of the scaling factors $S_1S_2$ which determines
unambiguously their scaling with the dihedral angle $\alpha$.
Using Eqs.\ (\ref{bound_alpha}) we get

\begin{equation}
\begin{array}{ccccl}
qp_1 & \sim & S_1^{-1/3}S_2^{2/3}S_1
& \sim & \alpha^{2/3} \\
(1/q)p_1'' & \sim & S_1^{1/3}S_2^{-2/3}
S_1S_2^2 & \sim & \alpha^{4/3} \\
\displaystyle{\int {d\tilde x \over q(\tilde x)}
\int {d\xi \over \xi^2}\left({d\phi \over d\xi}\right)^2}
& \sim & S_1^{1/3}S_2^{-2/3}S_1^2S_2^3 & \sim & \alpha^{7/3}.
\end{array}
\end{equation}

\noindent One can also predict how the width of the ridge 
which scales with $S_2^{-1}$ depends on the dihedral angle
$\alpha$ using one extra assumption.
If the transverse curvature decays exponentially away from the ridge
then the width of the ridge scales as the radius of curvature
times the bend angle $\alpha$.  Therefore

\begin{equation}
{\rm ridge\ width} \sim  S_2^{-1} \sim \alpha^{-1/3}.
\end{equation}

\noindent This equation fixes the dependence of the scale 
factors $S_1$ and $S_2$ on $\alpha$ so that
solutions for all dihedral angles can be generated from
a single solution for some $\alpha$.
All of the $\alpha$-scaling predictions have been convincingly verified
by the numerics.

\section{Numerical simulation of the stretching ridge}

We modeled an elastic sheet as a triangular lattice
of springs of unstretched length $a$ and spring constant $K$
after Seung and Nelson \cite{seung}.
Bending rigidity is introduced by assigning an energy
$J(1-\hat{\bf n}_1\cdot\hat{\bf n}_2)$ to every pair of 
adjacent lattice triangles with normals $\hat{\bf n}_1$ and 
$\hat{\bf n}_2$.  Seung and Nelson showed that when strains
are small and radii of curvature are large compared to the lattice
constant, this model membrane bends and stretches as a sheet
of thickness $h=a\sqrt{8J/K}$
made of isotropic homogeneous material with Young's modulus
$Y=2Ka/(h\sqrt{3})$ and Poisson ratio $\nu = 1/3$.
The bending modulus is $\kappa = J\sqrt{3}/2$.
We use a conjugate gradient energy minimization routine to
obtain a sequence of relaxed
shapes with the varying geometric parameter $\lambda$.
This lattice model has been used by the author
and others to verify the $\lambda$-scaling \cite{science}.
Here we explore the other two types of scaling which we
are convinced govern the ridge properties.

\begin{figure}
\centerline{\epsfxsize=8cm \epsfbox{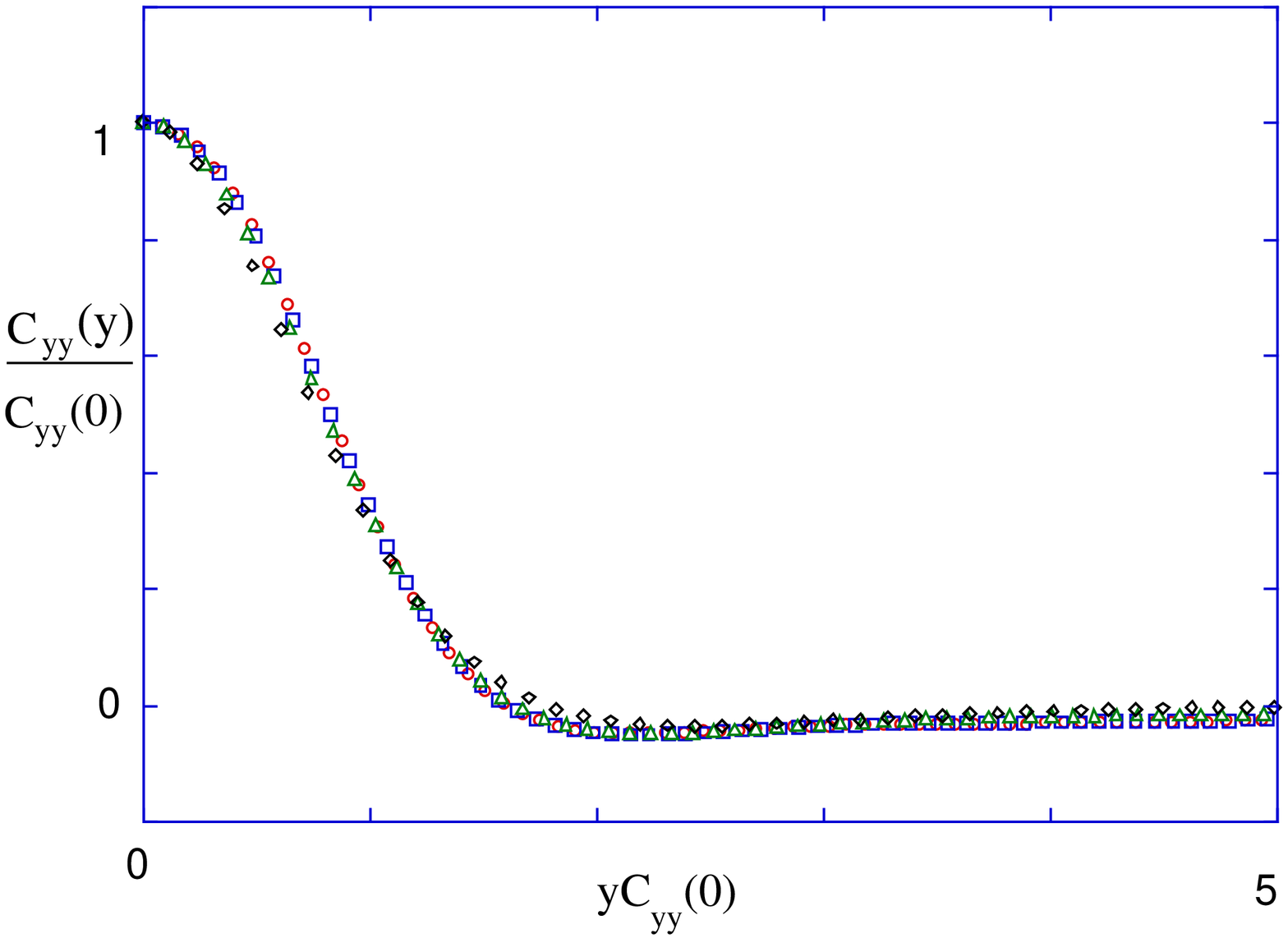}}
{FIG. 3. Transverse curvatures $C_{yy}(x,y)$
for $x=0$ (circles), $x=5$ (squares),
$x=10$ (triangles) and $x=15$ (diamonds) 
each scaled by $C_{yy}(x,0)$ vs.  the
transverse coordinate $y$ scaled by $C_{yy}^{-1}(x,0)$.
The curvatures are found numerically from a $50a$ by
$500a$ strip bent by a $90$ degree angle.}
\end{figure}

A rather suggestive demonstration of the $q(\tilde x)$-scaling
is presented on Fig.~3.  We plot the transverse curvatures
for different slices $\tilde x = const$ obtained from
a simulated strip of dimensions $50a$ by $500a$ bent
by normal forces applied to the $x=\pm X/2$ parts of the
boundary to make a $90$ degree angle.  The dimensionless thickness
is $\lambda \simeq 10^{-3}$.
The curvature for each slice is scaled by its value at the
origin $\tilde y = 0$.  The transverse coordinate is scaled by the
inverse of the curvature at $\tilde y = 0$.  These profiles
collapse onto a single scaling curve.
A more quantitative test is based on the prediction for the
variation of the transverse radius of curvature along the ridge 
({\it i.~e.~} at $\xi =0$)
$R_{yy} = 1/C_{yy} \sim q(\tilde x)$.
The coefficients $b_2$ and $b_4$ in the expansion
$R_{yy}(\tilde x) \simeq R_{yy}(0)(1 + b_2 \tilde x^2 + b_4 \tilde x^4)$
can be extracted from the numerics and compared to those
obtained from Eq.\ (\ref{scale_factor}).

Here we can also test the dependence of the scaling on the
boundary conditions which cause the ridge singularity.
Instead of using a long strip bent by normal forces applied to
the boundaries we can consider any shape which in the limit
of zero thickness exhibits a sharp crease.  One such shape
chosen due to ease of implementation is a regular tetrahedron
which was previously used to verify the $\lambda$-scaling \cite{science}.
A picture of one such minimum-energy tetrahedral surface is presented
in Fig.~4. Shading is proportional to the local stretching energy.

\begin{figure}
\centerline{\epsfxsize=8cm \epsfbox{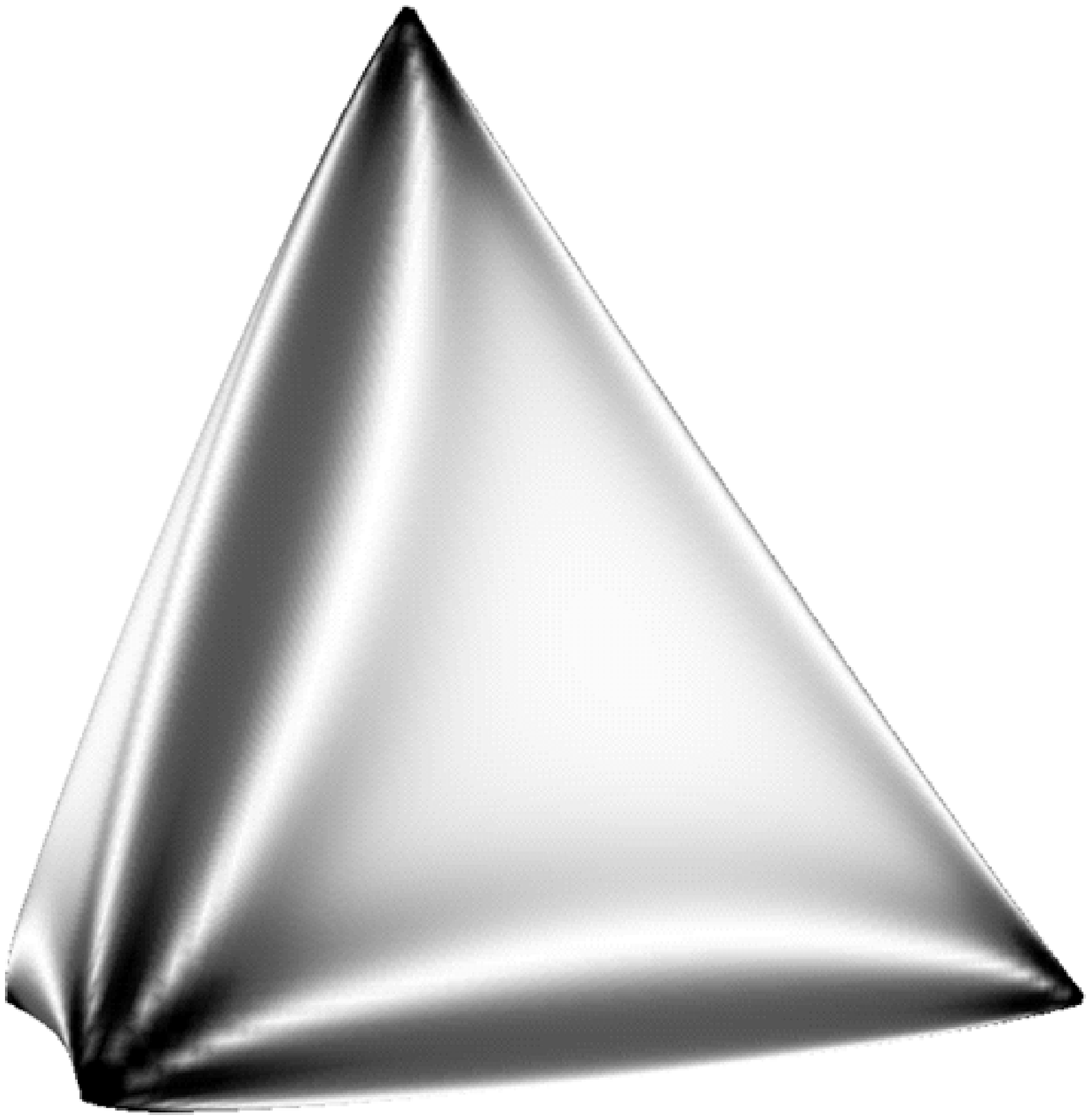}}
{FIG. 4. A tetrahedron of side $X=100a$ and thickness $h=0.063a$.
Shading is proportional to the local stretching energy.
Note the ``sagging'' of the ridge.}
\end{figure}

\begin{figure}
\centerline{\epsfxsize=8cm \epsfbox{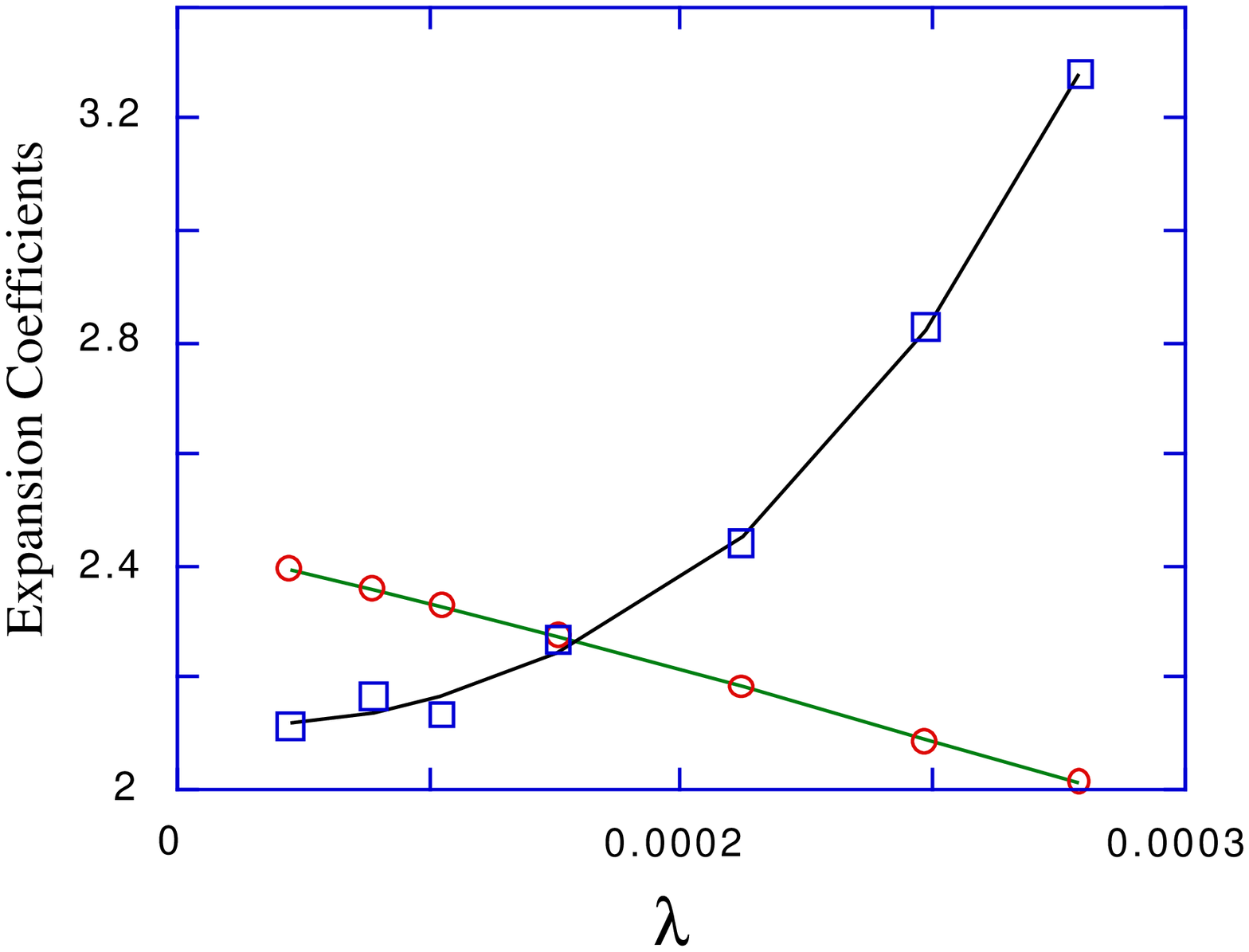}}
{FIG. 5. Scale factor expansion coefficients $b_2$ (circles)
and $b_4$ (squares) as extracted from a least squares fit to the
functional form of the transverse radius of curvature along the
ridge for tetrahedra of $X=100a$.}
\end{figure}

Figure 5 displays a plot of the coefficients $b_2$ (circles) 
and $b_4$ (squares) versus
the dimensionless thickness $\lambda$ obtained
by a least squares fit to the functional forms of $R_{yy}$ in the
range of $|\tilde x| \le 0.2$ for tetrahedra of varying
thicknesses but fixed edge length $X=100a$.
We see that these coefficients have a well defined limit as
$\lambda \rightarrow 0$ which is approached algebraically.
The extrapolations to $\lambda=0$ give 
$b_2=2.53\pm 0.04$ and $b_4=2.09\pm 0.07$
as compared to the prediced values of
$b_2 \simeq 2.47$ and $b_4 \simeq 2.03$.

We tested the dihedral angle scaling by making a long
rectangular strip of dimensions $50a$ by $200a$
and applying normal forces to its long boundaries so as to bend it
by an angle $\alpha$.  For all angles we fixed 
$\lambda \simeq 5\times 10^{-4}$.
The results are displayed in Figures 6 and 7.
Fig.~6 is a plot of the total elastic energy in units of
the bending modulus $\kappa$ {\sl vs.} the anticipated scaling
variable $\alpha^{7/3}$.  To a good precision the energy does indeed
exhibit the predicted scaling behavior.  The deviation at the
small bending angles is due to the finite size effects.
Since the width of the ridge diverges as $\alpha \rightarrow 0$
larger sheets are needed for smaller angles to avoid finite 
size effects.
In Fig.~7 the midridge curvature $C_{yy}(0)$
in units of $X^{-1}$ is plotted against 
$\alpha^{4/3}$.  The data agree well with the scaling prediction.

\begin{figure}
\centerline{\epsfxsize=8cm \epsfbox{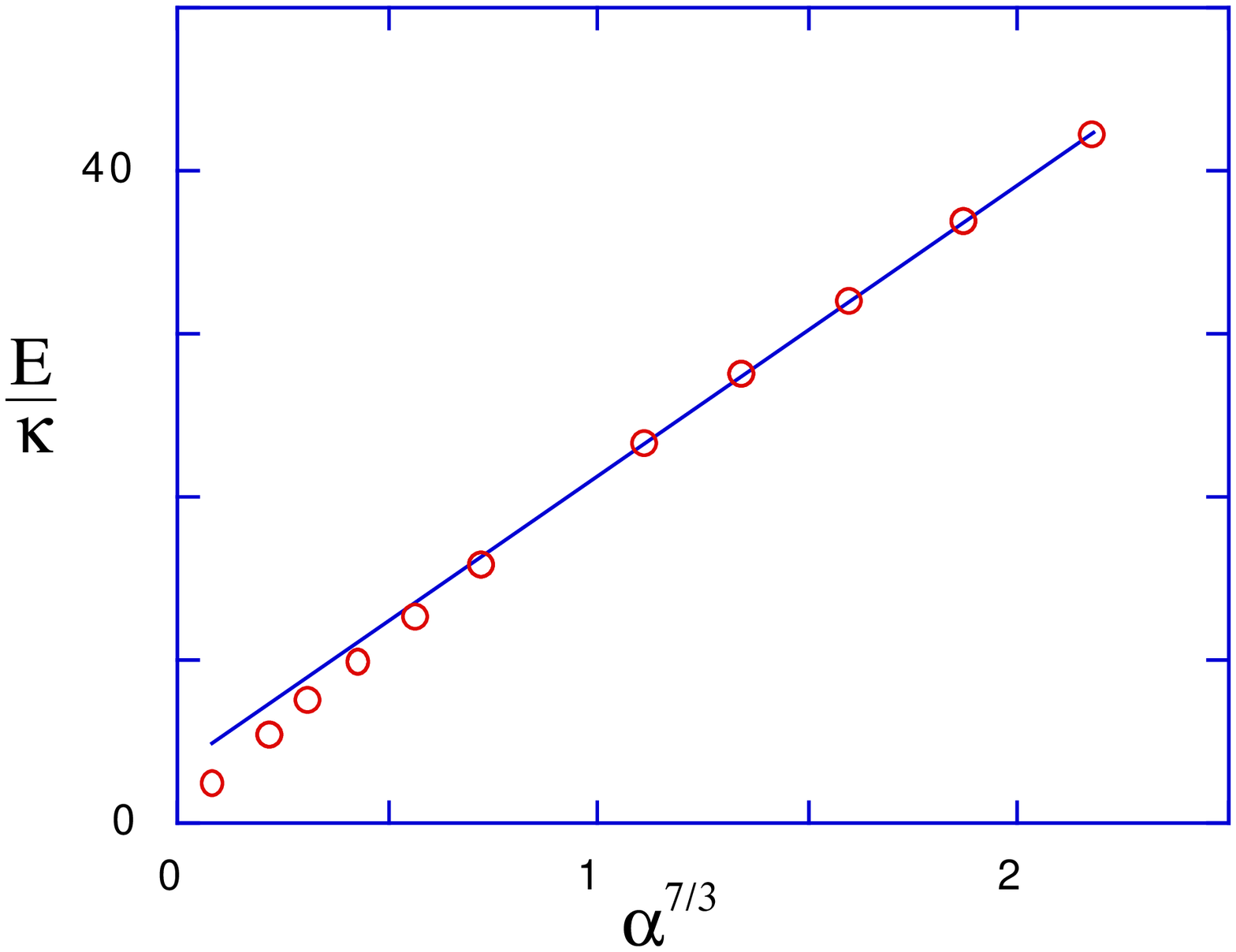}}
{FIG. 6. Total elastic energy in units of the bending
modulus $\kappa$ of $50a$ by $500a$ strips of thickness
$h = 0.063a$ bent by an angle $\alpha$.}
\end{figure}

Another test relevant to the question of the boundary condition 
dependence of the scaling predictions is the comparison of the
coefficients R for different boundary conditions in
the asymptotic form of the total elastic energy
of the ridge $E/\kappa \simeq R~ \lambda^{-1/3} \alpha^{7/3}$.
In a preceding work \cite{science} we found this coefficient for
the ridge appearing in the tetrahedral shape described above.
Its value, which was found by examining the dependence of the
energy on $\lambda$ for a fixed dihedral angle,
is $R_{tet} = 1.161 \pm 0.003$.  For the long strip
we have found $R_{strip} = 1.24 \pm  0.01$ by fixing $\lambda$
and varying $\alpha$.  Here the error range reflects only the
uncertainty arising from scatter in the numerical data.
However, there are additional errors resulting from
corrections to the asymptotic scaling not properly accounted
for in our crude fitting procedure.
We therefore suspect that the scaling coefficient $R$
depends insensitively on the boundary conditions and
varies only by a fraction of its value.  A stronger claim
of boundary condition independence of $R$ would require
a more convincing demonstration.

\begin{figure}
\centerline{\epsfxsize=8cm \epsfbox{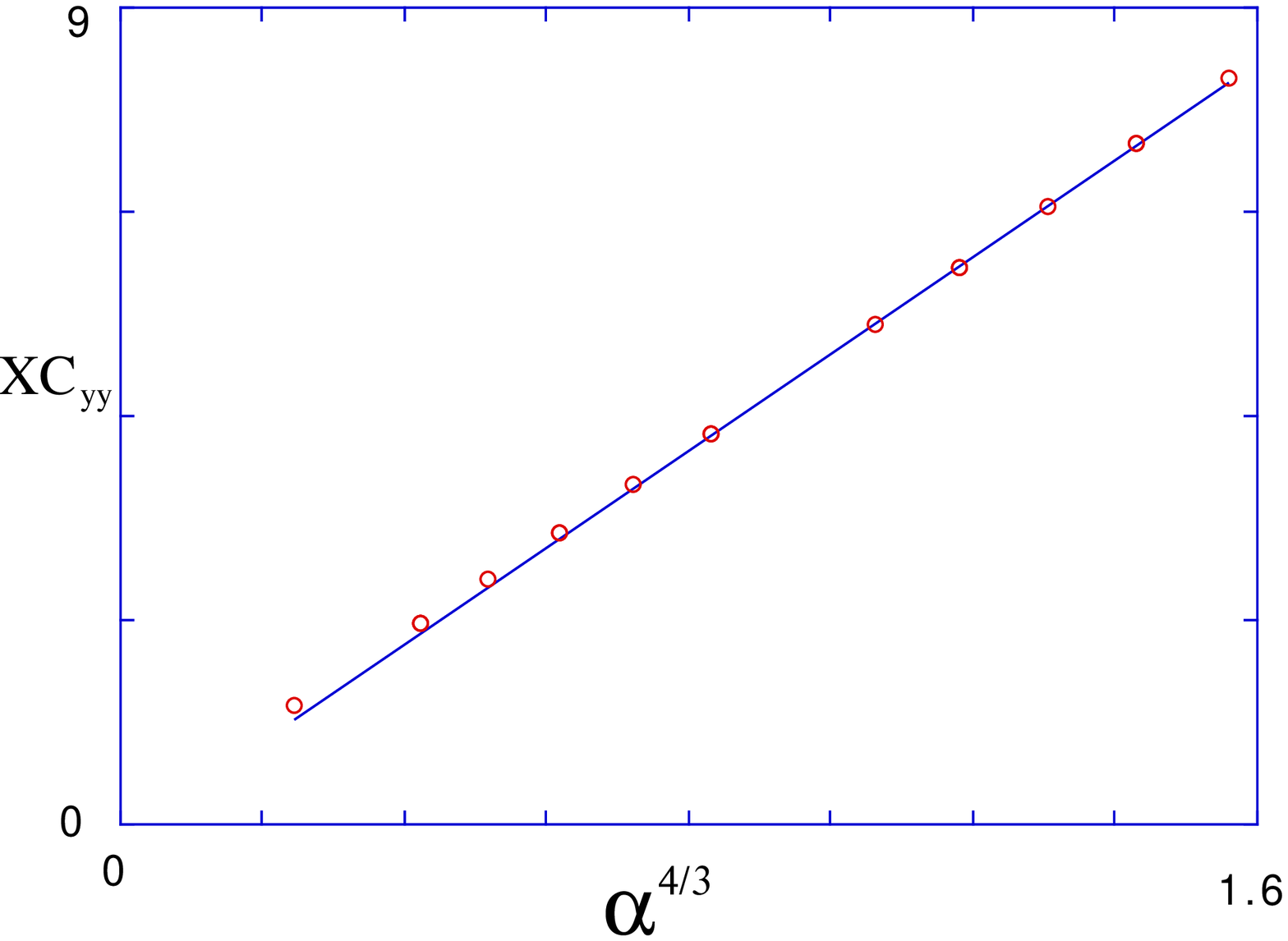}}
{FIG. 7. The midridge curvature in units of $X^{-1}$ 
for the $50a$ by $500a$ strips of thickness
$h = 0.063a$ vs. the anticipated scaling variable
$\alpha^{4/3}$.}
\end{figure}

\section{Discussion}

In this paper we described the ridge singularity
in thin elastic plates which arises under
certain types of boundary conditions.
We have found three types of scaling in the ridge
solution in the asymptotic
limit of small thickness of the elastic sheet.
The solution scales with the thickness of the
sheet in a non-trivial way.  In addition, it might scale
with the distance from the center of the ridge
and with the dihedral angle of the ridge.
To ascertain the importance of the results we ought to discuss
the degree to which the nature of the ridge singularity is 
independent of the boundary conditions.
First we note that the equations used to describe the
behavior of the elastic sheet assumed linear stress-strain
relations.  Since the strains in the ridge were
found to vanish in the limit of small thickness, the
results should be applicable to the real materials
in which the linear stress-strain relations are more
accurate at small strains.

The $\lambda$-scaling which was anticipated
from previous works\cite{fullerine,science}
has been put on a more rigorous footing.
It seems to be inevitable and
depends only on the geometry of the singularity.
It is by no means a unique type of scaling, since 
Scheidl and Troger \cite{scheidl} found that a ring
ridge which appears in strong buckling of a spherical
shell has a width which scales as $\lambda^{1/2}$ as opposed
to the $\lambda^{1/3}$ scaling for the straight ridge singularity
examined in this article.  The property that makes the
straight ridge important is that its energy grows slower
with size than that of the ring ridge which makes the
ring ridge unstable to breakup into straight ridges
when the cost of distorting the rest of the sphere
can be overcome by the energy gain from the formation
of the straight ridges.  We believe therefore that
straight ridges will be the dominating type of the
singularity in crumpled thin elastic sheets.
The morphology of a crumpled sheet can thus be
represented as a network of straight ridges.  
Since in the limit of small thickness the elastic energy
is concentrated into a progressively smaller fraction of
the crumpled sheet the ridges can at least in the
first approximation be treated as independent.
Therefore, once the ridge network is given characterization
in terms of the distribution of the lengths and the dihedral angles
of the ridges, the elastic energy of the crumpled sheet
is given by the sum
$E/\kappa \simeq R~\sum_i \lambda_i^{-1/3} \alpha_i^{7/3}$.

The second type of scaling found in the ridge singularity is
scaling with the distance from the center $x=0$.
By assuming this scaling we are led to a separable dependence
on the longitudinal and transverse position.  Neither
the scaling nor the separability, however, have been proven.
We have only found suggestive numerical evidence supporting
separability by examining different slices of the ridge as in Fig.~3.
In addition, the first two coefficients in the
Taylor expansion of the scale factor $q(\tilde x)$ have been
found to agree with the numerics.  A more rigorous test of 
separability is still needed.  The transverse profile of the
ridge is most probably dependent
on the boundary conditions.  In Ref.~\cite{science} the author
and others found that
the degree to which the transverse curvature oscillated away 
from the ridge in the tetrahedron shape is different from that
in the strip shape considered in this work.
The simplicity of the transverse
equations also needs to be better understood.
Usually such simplicity reflects some underlying
physical symmetry in the problem.

The third type of scaling exhibited by the ridge solution is 
the dihedral angle scaling.  This scaling
only depended on the {\it form} of the transverse
equations and not on the nature of their solution
and thus is likely to be independent of the
boundary conditions.  Numerical evidence in Fig.~6 and Fig.~7
supports this conclusion.  In addition,
we found the same dihedral angle scaling in the strip
shape as in the rhombus ``kite'' shape of Ref.~\cite{science}.

In future work we plan to analyze the transverse
equations and relate them to the boundary conditions in order to 
clarify whether they possess boundary condition independent features.
A study of how non-asymptotic effects affect the validity
of the separable solution is also in order.
The framework developed in this paper is suitable
to study of ridges under loading which is relevant to
mechanical properties of crumpled sheets. 
One could also introduce thermal shape fluctuations
to study how they affect the scaling properties of
the singularity, in particular, whether the
singular behavior is robust with respect to 
introduction of the fluctuations.

\section*{Acknowledgements}

Presented as a thesis to the Department of Physics, 
The University of Chiacgo, in partial fulfillment of the
requirements for the Ph.~D.~degree under the supervision
of Prof.~T.~A.~Witten.
This work was supported in part by the National Science
Foundation under the contract DMR-9208527 and by the MRSEC Program
of the National Science Foundation under the Award Number DMR-9400379.

\appendix
\section*{}

To show that Eqs.\ (\ref{zeroth}) only separate for one choice of the
scaling exponents $\mu$ and $\eta$ in the scaling Ansatz Eq.\ (\ref{Ansatz})
we carry out the substitution for general exponents.
We obtain

\begin{equation}
\begin{array}{rcl}
p_1'''' & = & q''q^{\eta+1}\left[\eta p_1''p_2 + \mu p_2'' p_1 
- \xi(p_1'' p_2' + p_2'' p_1')\right] + \\[1mm]
& & +(q')^2 q^\eta\left\{\eta(\eta-1)p_1''p_2 + \mu(\mu-1)p_2''p_1 - \right.
\\[1mm]
& & \left.-2(\mu-1)(\eta-1)p_1'p_2'\right\} \\[1mm]
p_2'''' & = & -q''q^{2\mu-\eta+1}\left[\mu p_1''p_1 -
\xi p_1''p_1'\right]+\\[1mm]
& & +(q')^2 q^{2\mu-\eta}\left\{(\mu-1)^2(p_1')^2 - 
\mu(\mu-1) p_1''p_1\right\}.
\end{array}
\end{equation}

\noindent The separation-of-variables conditions on the
scale function $q(\tilde x)$ demand that
$q''q^{\eta + 1} = A_1$ and $q''q^{2\mu-\eta+1} = A_2$ which requires
that $\mu = \eta$.
Unless the factors in curly braces vanish, separability also demands
that $(q')^2 q^\eta = A_3$ and $(q')^2 q^{2\mu-\eta} = A_4$.
From the $A_1$ and $A_3$ conditions it follows that
$q''q = (A_1/A_3) (q')^2$ which implies that $q' \propto q^{A_1/A_3}$.
Since we require that $q'(0) = 0$ and $q(0) \ne 0$, one of the following two
conditions must be true.  First $A_1$ may vanish which leads to
an unphysical choice $q'' \equiv 0$.  Second, the factors in
curly braces may vanish which happens only if $\mu=\eta=1$.
We conclude that indeed $\mu=\eta=1$ as asserted in the text.

\end{multicols}
\end{document}